\documentclass[aps,   groupedaddress, showkeys, showpacs]{revtex4}
\usepackage{amsmath,amsfonts,amscd,amssymb,epsf}

\newcommand{\Z}{{\mathbb{Z}}}

\newcommand{\R}{\mathbb{R}}

\newcommand{\pa}{\partial}

\begin{document}

\title
{Repulsive Casimir Force from  Fractional Neumann Boundary Conditions}

\author{S.C. Lim }  \author{L.P.
Teo }\address{Multimedia University, Jalan Multimedia, 63100, Cyberjaya, Selangor, Malaysia.}\email{sclim@mmu.edu.my; lpteo03@yahoo.com}


 \begin{abstract}
This paper studies the finite temperature Casimir force acting on a rectangular piston  associated with a massless fractional Klein-Gordon field at finite temperature. Dirichlet boundary conditions are imposed on the walls of a $d$-dimensional rectangular cavity, and a fractional Neumann condition is imposed on the piston that moves freely inside the cavity. The fractional Neumann condition gives an interpolation between the Dirichlet  and   Neumann conditions, where the Casimir force is known to be always attractive and always repulsive respectively. For the fractional Neumann boundary condition,  the attractive or repulsive nature of the Casimir force is governed by the fractional order which takes values from zero (Dirichlet) to one (Neumann). When the fractional order is larger than 1/2,  the Casimir force is always repulsive. For some fractional orders that are less than but close to 1/2, it is shown that the Casimir force can be either attractive or repulsive depending on the aspect ratio of the cavity and the  temperature.
\vspace{0.3cm}

\end{abstract}

\maketitle
\section{Introduction}
During the past decade, fractional calculus has attracted considerable attention from physicists and engineers. This is particularly true for researchers working in the fields of condensed matter physics, where fractional differential equations have been used to model various anomalous transport phenomena \cite{1,2,3,4,5}. Thus one expects fractional calculus, in particular fractional differential equations, to play an important role in quantum theories of mesoscopic systems and soft condensed matter which exhibit fractal character. Fractional dynamics provides a natural framework for describing the evolution of physical systems in fractal and multifractal media. By extending the above argument to quantum theories in fractal space-time, one may then have to deal with quantum mechanics and quantum field theory which satisfy fractional generalizations of Schr\"odinger, Klein-Gordon and Dirac equations. However, applications of fractional calculus to quantum theory are still relatively new. In quantum mechanics, fractional Schr\"odinger equation in its various forms has been studied recently \cite{6,7,8,9,10}. Fractional supersymmetric quantum mechanics has also been considered \cite{11}. Fractional Klein-Gordon equation \cite{11,12,13,14,15,16} and fractional Dirac equation \cite{17,18,19} have been considered more than a decade ago for various reasons. For examples, field theory with Lagrangian containing nonlocal kinetic terms involving fractional power of D'Alembertian operator arises in the (2 + 1)-dimensional bosonization \cite{20,21} and also in effective field theory with some degrees of freedom integrated out in the underlying local theory \cite{22,23}. Some studies have indicated that the QED radiative correction leads to a modification of the propagator of a charged Dirac particle which acquires a fractional exponent connected with the fine structure constant that is a fractal propagator \cite{24,25,26}. In quantum field theory of gravity with asymptotic safety \cite{27}, spacetime geometry cannot be understood in terms of a single metric. There is a need to introduce a different effective metric at each momentum scale. In addition, spacetime structure in asymptotically safe quantum Einstein gravity at sub-Planckian distances has been shown to be fractal \cite{28,29}. Numerical simulations have studied the propagation of a scalar particle in a dynamically triangulated spacetime, with a discretized version of the Einstein dynamics, and found that the spectral dimension of the microscopic spacetime is two and it tends to four for long time scales \cite{30,31,32}.  The use of differential and integral operators of fractional constant and variable orders may seem appropriate and natural for such quantum field theories in fractal and multifractal spacetime.

Recently, work on Casimir energy associated with fractional Klein-Gordon field has been carried out, in particular the Casimir effect associated with fractional massive and massless fields has been studied \cite{33,34}. Self-interacting scalar massive and massless fractional Klein-Gordon fields have been considered in the context of topological symmetry breaking on toroidal spacetime \cite{35}. In this paper, we want to study Casimir effect corresponding to fractional Klein-Gordon massless scalar field in the piston setting. Casimir effect associated with piston geometry has attracted considerable attention since it was first introduced by Cavalcanti \cite{36}. The geometric setup of Casimir piston  can be used to avoid divergence problems that usually plague the calculations of Casimir effect.  By taking suitable limits, the piston approach can be used to  derive the Casimir force acting on two parallel plates embedded orthogonally inside an infinitely long chamber \cite{37}, and the Casimir force acting on two infinite parallel plates \cite{38,39}.
Therefore it would be interesting to study the behavior of the Casimir force acting on a $d$-dimensional rectangular piston due to fractional Klein-Gordon massless field. In line with our consideration with the fractional field, we also impose on the piston fractional Neumann boundary condition which allows the interpolation between the ordinary Neumann boundary condition and the Dirichlet boundary condition.

\section{ Casimir piston for fractional Klein Gordon field with fractional Neumann conditions}

\begin{figure}\centering \epsfxsize=.35\linewidth
\epsffile{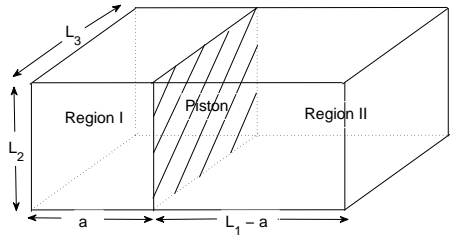}\caption{\label{f1} A three dimensional
  piston.}\end{figure}

Let $\phi(\mathbf{x},t)$, $\mathbf{x}\in\R^d, t\in \R$ be a massless fractional Klein-Gordon field with Lagrangian
\begin{equation*}
\mathcal{L}=\frac{1}{2}\phi(\mathbf{x},t)(-\Delta)^{\alpha}\phi(\mathbf{x},t),
\end{equation*}where $\Delta = \frac{\pa^2}{\pa t^2}+\sum_{i=1}^d \frac{\pa^2}{\pa x_i^2}.$ For the interpretation of the fractional operator $(-\Delta)^{\alpha}$, we refer to our previous work \cite{34}. The purpose of this paper is to investigate the finite temperature Casimir force acting on a piston moving freely inside a rectangular cavity $[0,L_1]\times\ldots\times[0, L_d]$ (see Figure \ref{f1}) due to the fractional field $\phi(\mathbf{x},t)$. We assume that the piston   has negligible thickness and is orthogonal to the $x_1$ direction, with position given by $x_1=a$, where $0<a<L_1$. On the walls of the rectangular cavity, the field $\phi(\mathbf{x},t)$ is assumed to satisfy Dirichlet boundary conditions, i.e.
$\left.\phi(\mathbf{x},t)\right|_{x_i=0\;\text{or}\;L_i}=0,  i=1,\ldots, d$. On the piston, the field $\phi(\mathbf{x}, t)$ is assumed to satisfy the fractional Neumann condition
\begin{equation}\label{eq5_19_1}
\left.\frac{\pa^{\eta}}{\pa x_1^{\eta}}\phi(x_1, x_2,\ldots, x_d,t)\right|_{x_1=a}=0.
\end{equation}The cases where $\eta=0$ (Dirichlet boundary condition) and $\eta=1$ (Neumann boundary condition) have been discussed in our previous work \cite{5_19_2}. Therefore in this paper, we only consider the case where $\eta\in (0,1)$.

Assume that the system is maintained in thermal equilibrium at temperature $T$. The Casimir free energy of the system is the sum of the Casimir free energy $E_{\text{Cas}}^{\text{I}}$ inside Region I $[0, a]\times[0, L_2]\times\ldots\times[0, L_d]$, the Casimir  free energy $E_{\text{Cas}}^{\text{II}}$ inside Region II $[a, L_1]\times[0,L_2]\times\ldots\times[0,L_d]$, and the Casimir free energy $E_{\text{Cas}}^{\text{out}}$ outside the cavity $[0, L_1]\times\ldots\times[0, L_d]$, i.e.,
\begin{equation*}
E_{\text{Cas}}^{\text{piston}}= E_{\text{Cas}}^{\text{I}}+E_{\text{Cas}}^{\text{II}}+E_{\text{Cas}}^{\text{out}}.
\end{equation*}The Casimir free energy of the exterior region $E_{\text{Cas}}^{\text{out}}$ does not give rise to Casimir force on the piston. Using Matsubara formalism, the Casimir energy inside the cavity $[0, a]\times[0, L_2]\times\ldots\times[0, L_d]$ is given by
\begin{equation*}
E_{\text{Cas}}^{\text{I}}(a)= \frac{T}{2}\log \det \left(\left[-\frac{\Delta}{\mu^2}\right]^{\alpha}\right),
\end{equation*}where $\mu$ is a normalization constant with dimension length$^{-1}$, $\det ([-\Delta/\mu^2]^{\alpha})$ is the determinant of the fractional Klein-Gordon operator $[-\Delta/\mu^2]^{\alpha}$ acting on functions $\phi(\mathbf{x}, t)$ on $[0,a]\times[0, L_2]\times\ldots\times[0, L_d]\times [0, 1/T]$, which satisfy Dirichlet boundary conditions on the boundaries $x_i=0$, $1\leq i\leq d$, $x_i=L_i$, $2\leq i\leq d$, satisfy fractional Neumann condition \eqref{eq5_19_1} on the boundary $x_1=a$ and satisfy periodic boundary condition in the $t$ direction, i.e.,
$
\phi\left(\mathbf{x}, t+\frac{1}{T}\right)=\phi\left(\mathbf{x}, t\right).
$ As is discussed in \cite{34}, a complete set of eigenfunctions is given by
\begin{equation*}
\begin{split}
\phi_{\boldsymbol{k}, l}(\mathbf{x}, t)=\sin\left(\frac{\pi\left(k_1-\frac{\eta}{2}\right)x_1}{a}\right)\prod_{i=2}^d\sin \left(\frac{\pi k_i x_i}{L_i}\right) e^{2\pi i l T t},
\end{split}
\end{equation*}with eigenvalues $\left[\omega_{k,l}^2/\mu^2\right]^{\alpha}$, where
\begin{equation*}
 \omega_{\boldsymbol{k}, l}^2 = \left( \frac{\pi\left(k_1-\frac{\eta}{2}\right)}{a}\right)^2 +\sum_{i=2}^{\infty}\left(\frac{\pi k_i }{L_i}\right)^2+ (2\pi l T)^2.
\end{equation*}Here $\boldsymbol{k}=(k_1,k_2,\ldots, k_d)$, $k_1\in \Z$, $k_2, \ldots, k_d\in \mathbb{N}$ and $l\in \Z$. The Casimir energy $E_{\text{Cas}}^{\text{I}}(a)$ is then given by
\begin{equation*}
\begin{split}
E_{\text{Cas}}^{\text{I}}(a)=\frac{\alpha T}{2} \sum_{\boldsymbol{k}}\sum_{l} \log \frac{\omega_{\boldsymbol{k}, l}^2}{\mu^2}.
\end{split}
\end{equation*}We   regularize this sum by exponentially cut-off method, i.e.,
\begin{equation}\label{eq6_2_1}
\begin{split}
E_{\text{Cas}}^{\text{I}}(a;\lambda)=\frac{\alpha T}{2} \sum_{\boldsymbol{k}}\sum_{l}\left( \log \frac{\omega_{\boldsymbol{k}, l}^2}{\mu^2}\right)e^{-\lambda \omega_{\boldsymbol{k}, l}},
\end{split}
\end{equation}where $\lambda$ is a cut-off parameter with the dimension of length.
For the Casimir energy in Region II, one can show that it can be obtained from the Casimir energy in Region I \eqref{eq6_2_1} by replacing $a$ with $L_1-a$.
Therefore, using the result    \eqref{eq5_20_1} in the Appendix, we find that up to the term constant in $\lambda$,
\begin{equation}\label{eq5_20_8}
\begin{split}
&E_{\text{Cas}}^{\text{I}}(a;\lambda)+E_{\text{Cas}}^{\text{II}}(a;\lambda)=-\alpha T\sum_{i=2}^{d+1}\frac{\Gamma(i)}{\Gamma\left(\frac{i}{2}\right)}c_{d+1-i}(L_1)\left(\log[\lambda\mu]^2-2\psi(i)\right)\lambda^{-i} -\frac{\alpha T}{2}\Bigl(\zeta'(0;a)
+\zeta'(0; L_1-a)\Bigr),
\end{split}
\end{equation}where
\begin{equation*}\begin{split}
c_i(L_1)=(-1)^i\frac{L_1 S_{d-1-i}}{2^d\pi^{\frac{d+1-i}{2}}T}, \hspace{1cm}  S_{d-1-i}:=\sum_{2\leq j_1<j_2<\ldots<j_{d-1-i}\leq d} L_{j_1} \ldots L_{j_{d-1-i}},\end{split}
\end{equation*}  and $\zeta(s; a)$ is the zeta function
\begin{equation*}
\begin{split}
\zeta(s;a)= \sum_{\boldsymbol{k}\in \Z\times\mathbb{N}^{d-1}}\sum_{l=-\infty}^{\infty}\omega_{\boldsymbol{k}, l}^{-2s}.
\end{split}
\end{equation*}
The first summation in \eqref{eq5_20_8} is the sum of the $\lambda\rightarrow 0^+$ divergent terms. We notice that these divergent terms  are independent of the position of the piston $a$ and the fractional parameter  $\eta$. The last term in \eqref{eq5_20_8} is what one would obtain for the Casimir energy if one uses the zeta regularization method to compute the Casimir energy.


Upon differentiating \eqref{eq5_20_8} with respect to $a$, we find that all the terms that diverge when $\lambda\rightarrow 0^+$ do not contribute to the Casimir force acting on the piston, since these terms are independent of $a$. By taking the limit $\lambda\rightarrow 0^+$, we find that the Casimir force acting on the piston is
\begin{equation}\label{eq5_20_4}\begin{split}
F_{\text{Cas}}(a;L_1)=&-\lim_{\lambda\rightarrow 0^+} \frac{\pa}{\pa a}\Bigl\{E_{\text{Cas}}^{\text{I}}(a;\lambda)+E_{\text{Cas}}^{\text{II}}(a;\lambda)\Bigr\}=\frac{\alpha T}{2}\frac{\pa}{\pa a}\Bigl(\zeta'(0;a)
+\zeta'(0; L_1-a)\Bigr).\end{split}
\end{equation}Using the result \eqref{eq5_20_5} in the Appendix, we find that the Casimir force can be written as the difference
\begin{equation*}
F_{\text{Cas}}(a;L_1)=F_{\text{Cas}}^{\infty}(a)-F_{\text{Cas}}^{\infty}(L_1-a),
\end{equation*}where
\begin{equation}\label{eq5_20_6}\begin{split}
&F_{\text{Cas}}^{\infty}(a) = -2\alpha T\text{Re}\; \sum_{(k_2,\ldots, k_d)\in\mathbb{N}^{d-1}}\sum_{l=- \infty}^{\infty}  \frac{e^{\pi i \eta}\sqrt{\sum_{i=2}^d\left(\frac{\pi k_i}{L_i}\right)^2+(2\pi l T)^2}}{\exp\left(2a \sqrt{\sum_{i=2}^d\left(\frac{\pi k_i}{L_i}\right)^2+(2\pi l T)^2}\right)-e^{\pi i \eta}} \end{split}
\end{equation} is the Casimir force acting on the piston when the right end of the cavity is   infinite distance away, i.e., $L_1\rightarrow \infty$. It can also be interpreted as the Casimir force acting between two parallel plates embedded inside an infinitely long rectangular cylinder, where   Dirichlet boundary condition is imposed on one of the plates, and fractional Neumann boundary condition is imposed on the other. It should be remarked that by putting $\eta=0$ and $\eta=1$ in \eqref{eq5_20_6}, we obtain respectively   twice the Casimir force when Dirichlet and Neumann boundary conditions are imposed on the piston. The factor of two arises since for Dirichlet and Neumann boundary conditions, the $k_1$ and $-k_1$ modes are dependent, but for general fractional Neumann conditions with $\eta\in(0,1)$, the $k_1$ and $-k_1$ modes are independent.

Since
\begin{equation}\label{eq5_20_7}
f(x):=\text{Re}\; \left\{\frac{e^{\pi i\eta}}{x-e^{\pi i \eta}}\right\}=\frac{x\cos(\pi \eta)-1}{(x-\cos(\pi \eta))^2+\sin^2(\pi \eta)},
\end{equation}we can conclude from \eqref{eq5_20_6} that the Casimir force is always repulsive when $\eta\geq 1/2$.  For $\eta\in (0,1/2)$, the Casimir force can be attractive or repulsive depending on the values of $a,   L_2, \ldots, L_d$ and $T$.  However, when $a$ is large enough, or more precisely, when
$$ a> \frac{-\log \cos(\pi\eta)}{2\sqrt{\sum_{i=2}^d \left(\frac{\pi }{L_i}\right)^2}},$$the Casimir force is always attractive.
Taking the derivative of $f(x)$ \eqref{eq5_20_7} with respect to $x$, we have
\begin{equation*}
f'(x)= \frac{-x^2\cos (\pi \eta)+2x-\cos(\pi \eta)}{\left((x-\cos(\pi \eta))^2+\sin^2(\pi \eta)\right)^2}.
\end{equation*}From this, we can deduce that when $\eta\geq 1/2$, the magnitude of the Casimir force is always a decreasing function of $a$.

From \eqref{eq5_20_6}, we can also deduce that in the  high temperature limit, the Casimir force grows linearly with temperature, with leading term  given by the sum of the terms with $l=0$.
In the low temperature limit, the Casimir force can be written as a sum of the zero temperature Casimir force and the temperature correction term, i.e.,
\begin{equation}\label{eq5_27_1}
\begin{split}
F_{\text{Cas}}^{\infty}(a) = F_{\text{Cas}}^{\infty}(a; T=0)+\Delta_T F_{\text{Cas}}^{\infty}(a),
\end{split}
\end{equation}where the zero temperature Casimir force is given by
\begin{equation*}
\begin{split}
F_{\text{Cas}}^{\infty}(a; T=0)  =&-\frac{\alpha}{\pi a}\sum_{\boldsymbol{k}\in \mathbb{N}^d}\cos\left(\pi k_1\eta\right)\frac{\sqrt{\sum_{i=2}^d\left(\frac{\pi k_i}{L_i}\right)^2}}{k_1} K_1\left(2k_1a \sqrt{\sum_{i=2}^d\left(\frac{\pi k_i}{L_i}\right)^2}\right)\\&-\frac{2\alpha}{\pi }\sum_{\boldsymbol{k}\in \mathbb{N}^d}\cos\left(\pi k_1\eta\right) \left(\sum_{i=2}^d\left(\frac{\pi k_i}{L_i}\right)^2\right)K_0\left(2k_1a \sqrt{\sum_{i=2}^d\left(\frac{\pi k_i}{L_i}\right)^2}\right),
\end{split}
\end{equation*}and the temperature correction is
\begin{equation*}
\begin{split}
& \Delta_T F_{\text{Cas}}^{\infty}(a)=-\frac{2\alpha T}{\pi} \sum_{(k_2,\ldots, k_d)\in \mathbb{N}^{d-1} }\sum_{l=1}^{\infty}\frac{\sqrt{\sum_{i=2}^d\left(\frac{\pi k_i}{L_i}\right)^2}}{l}  K_1\left(\frac{l}{T}\sqrt{\sum_{i=2}^d\left(\frac{\pi k_i}{L_i}\right)^2}\right)
\\&+\frac{\pi^2\alpha}{a^3}\sum_{k_1=-\infty}^{\infty}\sum_{(k_2,\ldots, k_d)\in \mathbb{N}^{d-1} }\frac{\left(k_1-\frac{\eta}{2}\right)^2}{\sqrt{\left(\frac{\pi\left(k_1-\frac{\eta}{2}\right)}{a}\right)^2+\sum_{i=2}^d\left(\frac{\pi k_i}{L_i}\right)^2}}  \frac{1}{\exp\left(\frac{1}{T}\sqrt{\left(\frac{\pi\left(k_1-\frac{\eta}{2}\right)}{a}\right)^2+\sum_{i=2}^d\left(\frac{\pi k_i}{L_i}\right)^2}\right)-1}.\end{split}
\end{equation*}Observe that the temperature correction term goes to zero exponentially fast when the temperature approaches zero.

The expression \eqref{eq5_20_6} for the Casimir force is suitable for studying the behavior of the Casimir force when $a\gg L_i$, $i=2,\ldots,d$ and $a\gg 1/T$. On the other hand, the expression \eqref{eq5_27_1} is suitable for studying the behavior of the Casimir force when $L_i\ll a\ll 1/T$, $i=2,\ldots,d$. To study the behavior of the Casimir force when $a\ll 1/T \ll L_i$ or $1/T\ll a \ll L_i$, $i=2,\ldots, d$, we need alternative expressions. The derivation of these alternative expressions is a little bit tedious, but can be done using the same techniques as in \cite{34} and \cite{5_19_2}. We list these alternative expressions for the Casimir force in the Appendix. From \eqref{eq5_27_3}, we find that when $a\ll 1/T \ll L_i$, $i=2,\ldots, d$, the leading terms of the Casimir force $F_{\text{Cas}}(a)$ are given by
\begin{equation}\label{eq5_27_6}
\begin{split}
 F^{\infty}_{\text{Cas}}(a)\sim &- \frac{\alpha}{2^{d}}\sum_{j=0}^{d-1}(-1)^{d-1-j} \frac{j+1}{\pi^{\frac{j+2}{2}} a^{j+2}} S_j \Gamma\left(\frac{j+2}{2}\right) \sum_{k_1=1}^{\infty}\frac{\cos\left(\pi k_1\eta\right)}{k_1^{j+2}} + \alpha  T E\left(-\frac{1}{2}; \frac{ \pi}{L_2},\ldots,\frac{\pi}{L_d}\right)\\&- \alpha T \sum_{j=0}^{d-1}  \frac{(-1)^{d-1-j} }{2^{d-j-2}\pi^{\frac{j+2}{2}}}S_j
  \Gamma\left(\frac{j+2}{2}\right)\zeta_R(j+2)T^{j+2}.
\end{split}\end{equation}  The remaining terms goes to zero exponentially fast. The dominating term of the Casimir force when $a\ll 1/T \ll L_i$ is thus \begin{equation}\label{eq5_27_7}
\begin{split}
 F^{\infty}_{\text{Cas}}(a)\sim &- \frac{\alpha}{2^{d}}  \frac{d \left[\prod_{i=2}^d L_i\right]}{\pi^{\frac{d+1}{2}} a^{d+1}}  \Gamma\left(\frac{d+1}{2}\right)\sum_{k_1=1}^{\infty}\frac{\cos\left(\pi k_1\eta\right)}{k_1^{d+1}}.
\end{split}\end{equation}When $1/T\ll a \ll L_i$, $i=2,\ldots, d$, we find from \eqref{eq5_27_4} that the leading terms of the Casimir force $F_{\text{Cas}}(a)$ are given by
\begin{equation}\label{eq5_27_8}
\begin{split}
F^{\infty}_{\text{Cas}}(a)=
-\frac{\alpha T}{2^{d-1}}\sum_{j=0}^{d-1}   \frac{(-1)^{d-1-j}}{\pi^{\frac{j+1}{2}}}\frac{j S_j }{a^{j+1}}\Gamma\left(\frac{j+1}{2}\right) \sum_{k_1=1}^{\infty}\frac{\cos(\pi k_1\eta)}{k_1^{j+1}}+ \alpha  T E\left(-\frac{1}{2}; \frac{ \pi}{L_2},\ldots,\frac{\pi}{L_d}\right).
\end{split}
\end{equation}The remaining terms goes to zero exponentially fast. The dominating term of the Casimir force when $  1/T \ll a\ll L_i$ is thus
\begin{equation}\label{eq5_27_9}
\begin{split}
F^{\infty}_{\text{Cas}}(a)=&
-\frac{\alpha T}{2^{d-1}} \frac{(d-1) \left[\prod_{i=2}^d L_i\right] }{\pi^{\frac{d}{2}} a^{d}}\Gamma\left(\frac{d}{2}\right)\sum_{k_1=1}^{\infty}\frac{\cos(\pi k_1\eta)}{k_1^{d}}.
\end{split}
\end{equation}

\begin{table}\caption{\label{t1}The value of $x_n$ where $B_n(x)=0$.}
\begin{tabular}{||c|c||c|c||c|c||}
\hline
\hline
$n$ & $x_n$ & $n$ & $x_n$& $n$ & $x_n$\\
\hline
1 & 1/3=0.33333333 & 2 & 0.422649731 & 3& 0.46165930\\
  4&0.48067038 & 5 & 0.49023768 & 6& 0.49508143\\
7 & 0.49752864& 8 & 0.49876076 & 9& 0.49937940\\
\hline
\hline
\end{tabular}\end{table}

From \eqref{eq5_27_7} and \eqref{eq5_27_9}, we find that the signs of the dominating terms of the Casimir force when $a\ll 1/T \ll L_i$ and when $  1/T \ll a\ll L_i$ are governed by the sign of the function
\begin{equation*}
\begin{split}
B_n(x)=\sum_{k=1}^{\infty}\frac{\cos (\pi k x)}{k^n},\hspace{0.5cm}x\in [0,1],
\end{split}
\end{equation*}when $n=d+1$ and $n=d$ respectively. When $n$ is even, $B_n(x)$ is up to a constant, the Bernoulli polynomial of degree $2n$. For any integer $n$, the function $B_n(x)$ has been studied in \cite{34}. It was shown that this function is decreasing  and has a unique zero in the interval $  [0,1]$. Moreover, this zero point is less that $1/2$. The point $x_n$ where $B_n(x)=0$ for $n=1,2,\ldots,9$ is given in Table \ref{t1}. One can observe that $x_n$ is an increasing function of $n$. Therefore, we can conclude that when $L_i\gg a, 1/T$,   $i=2,\ldots, d$, and $x_d < \eta < x_{d+1}$,  increasing the temperature tends to change the Casimir force from attractive to repulsive.
\begin{figure}\centering \epsfxsize=.4\linewidth
\epsffile{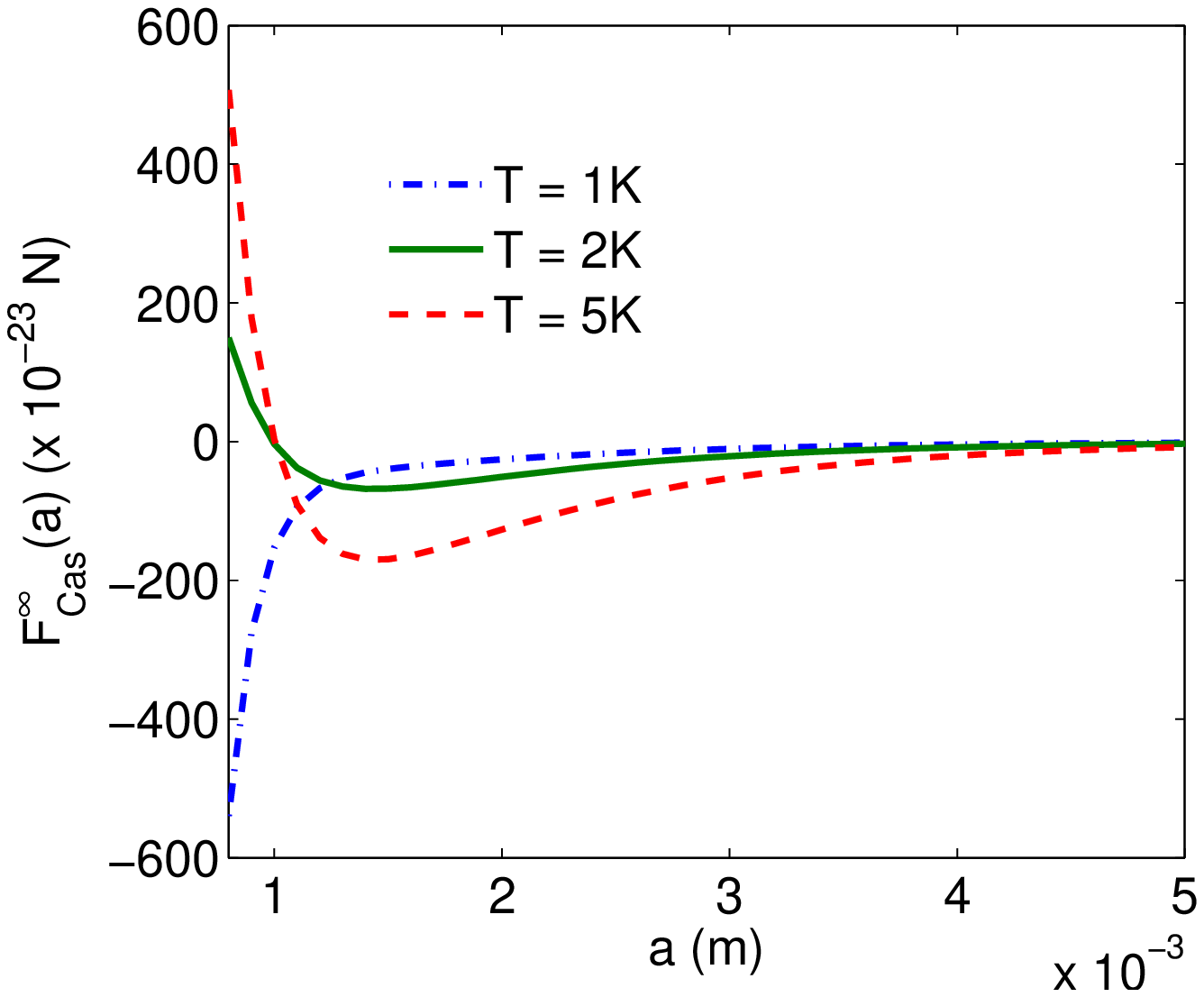}\centering \epsfxsize=.4\linewidth
\epsffile{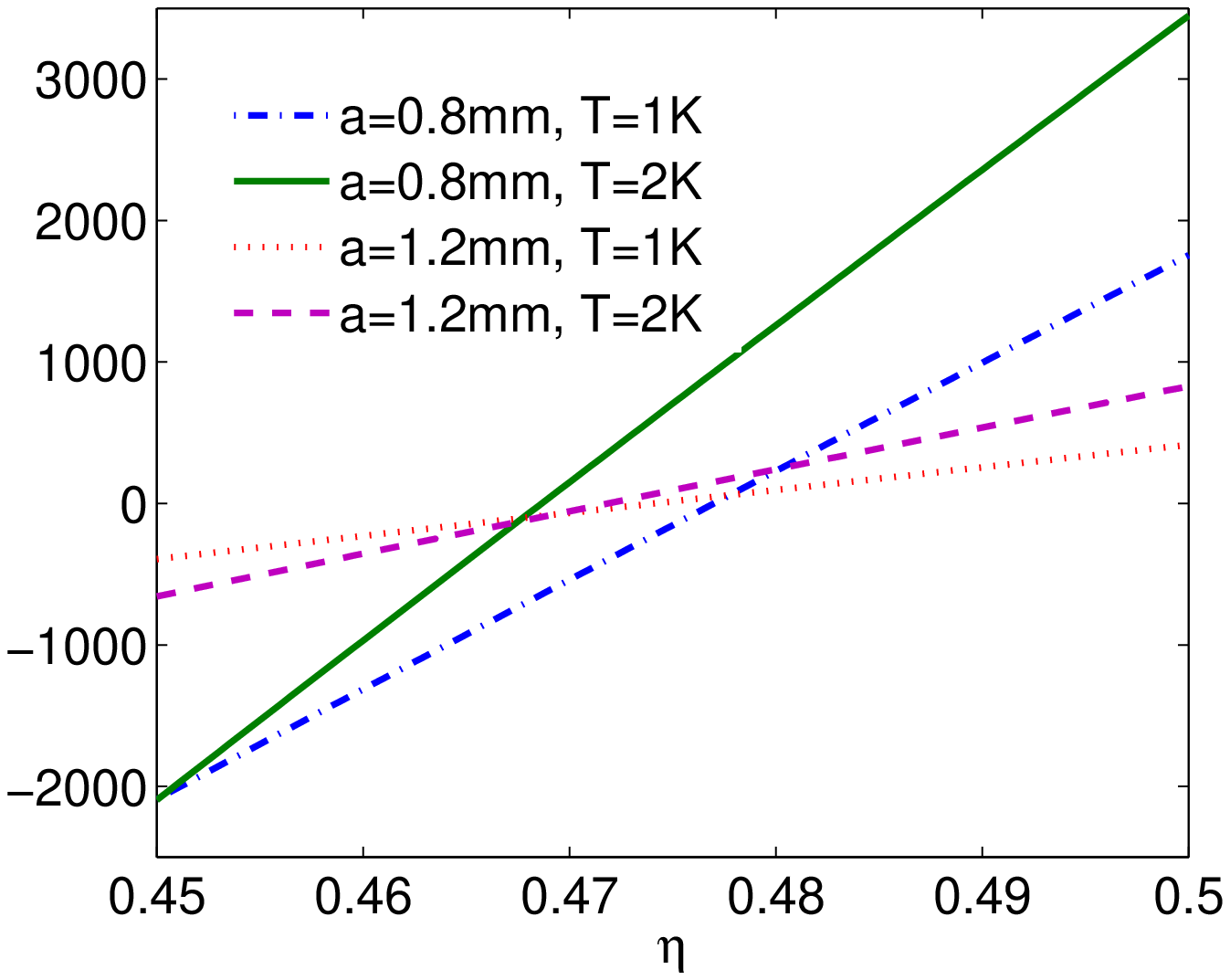}\caption{\label{f2} Left: The Casimir force $F_{\text{Cas}}^{\infty}(a)$ as a function of $a$ when $d=3$, $L_2=L_3=0.01$m, $\eta=0.47$, and $T= 1$K, 2K and 5K respectively. Right: The Casimir force $F_{\text{Cas}}^{\infty}(a)$ as a function of $\eta$ when $d=3$, $L_2=L_3=0.01$m for various values of $(a, T)$.}\end{figure}

\begin{figure}\centering \epsfxsize=.4\linewidth
\epsffile{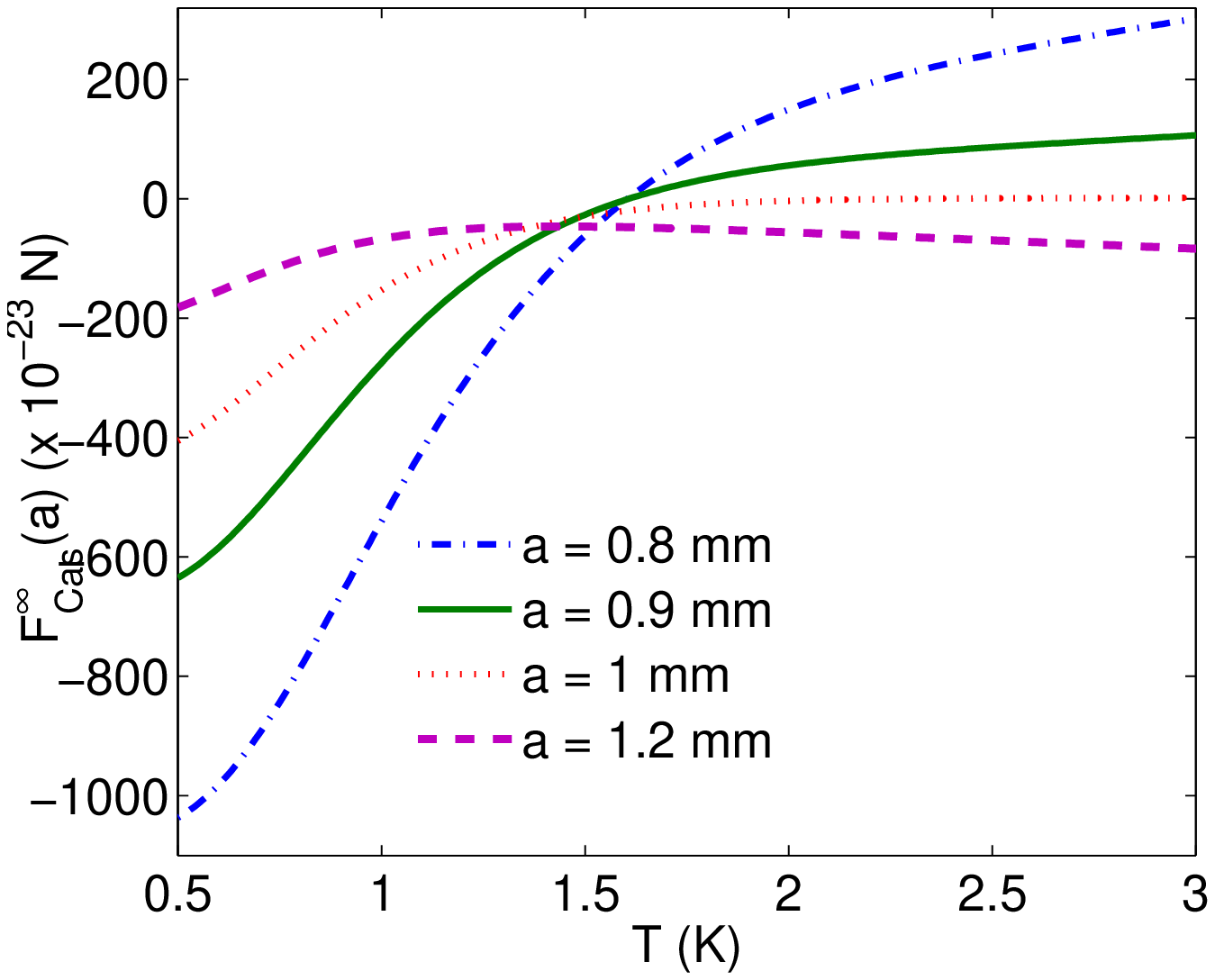}\centering \epsfxsize=.4\linewidth
\epsffile{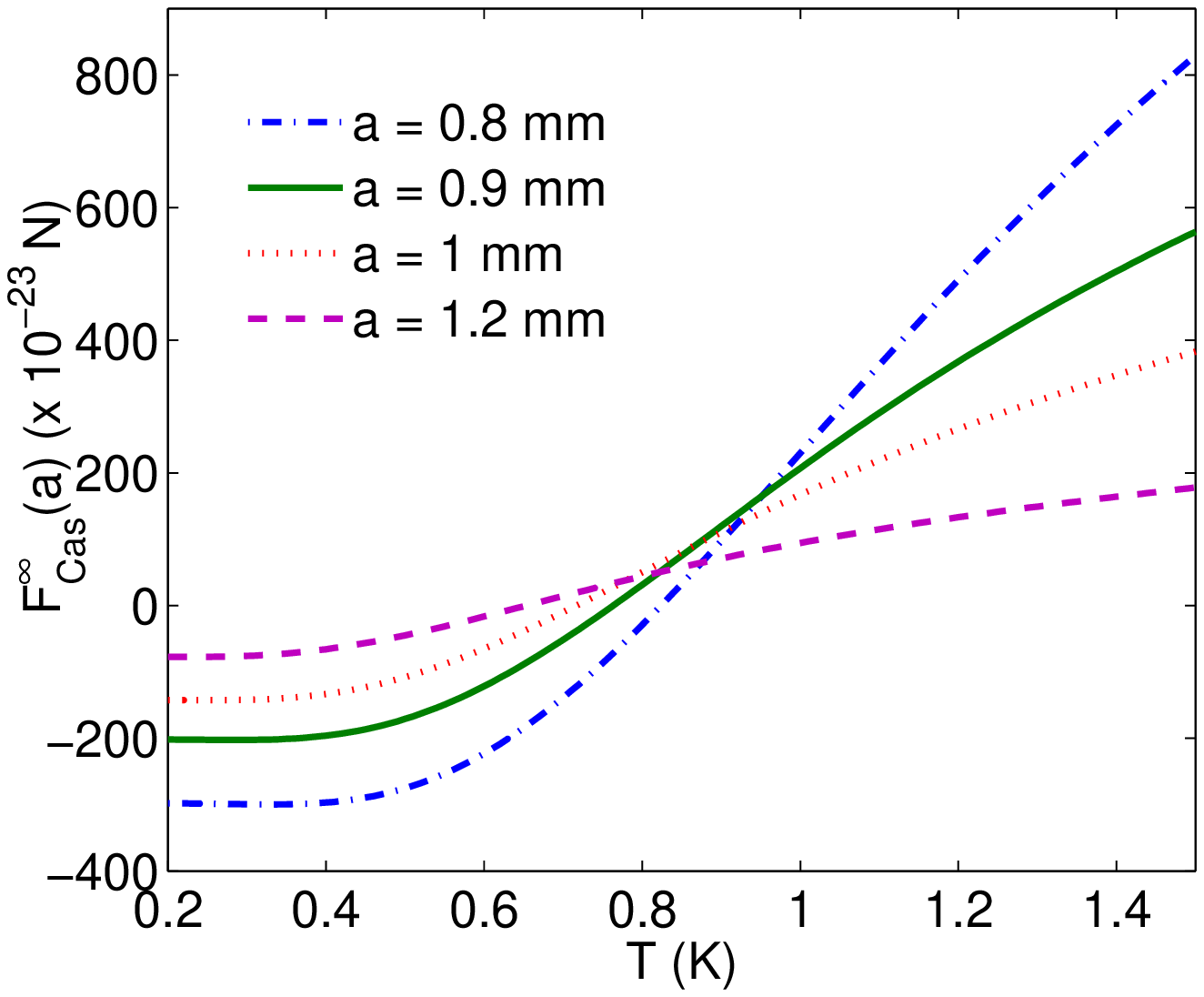}\caption{\label{f3} The Casimir force $F_{\text{Cas}}^{\infty}(a)$ as a function of $T$ when $d=3$, $L_2=L_3=0.01$m, $\eta=0.47$ (Left) and $\eta=0.48$ (Right), and $a= 0.8$mm, 0.9mm, 1mm and 1.2mm respectively.}\end{figure}

In FIG. \ref{f2}  and FIG. \ref{f3}, we show graphically the dependence of the Casimir force on various parameters such as $a$, $T$ and $\eta$ when the dimension $d$ is three. The figure on the left of FIG. \ref{f2} shows that when the fractional order $\eta$ is equal to $0.47$ and the temperature $T$ is equal to 2K or 5K, the Casimir force can change from repulsive to attractive when we increase $a$. The figure on the right of FIG. \ref{f2} shows the dependence of the Casimir force on the fractional order $\eta$. In general, the Casimir force changes from attractive to repulsive when we increase $\eta$. FIG. \ref{f3}  shows that when $\eta=0.47$ and $\eta=0.48$, increasing temperature can change the Casimir force from attractive to repulsive.

As a conclusion, we have studied the finite temperature Casimir effect on a piston moving freely inside a rectangular cavity due to a massless fractional Klein-Gordon field, where Dirichlet boundary conditions are imposed on the walls of the cavity and fractional Neumann boundary condition of order $\eta$ is imposed on the piston. We show analytically that for $\eta$ larger than $1/2$, the Casimir force is always repulsive. Moreover, the magnitude of the Casimir force is a decreasing function of $a$. This is another instance where we can obtain repulsive Casimir force in the piston setting besides the hybrid boundary conditions considered in \cite{6_2_3} and the Robin boundary conditions considered in \cite{6_2_2}.    The more interesting case is when $\eta$ is less than $1/2$. We show that for the values of $\eta$ less than $1/2$, the Casimir force can always become attractive when  $a$ is large enough. For certain values of $\eta$ less than but close to $1/2$, it has been demonstrated graphically that the Casimir force can be attractive or repulsive depending on the aspect ratio of the cavity and the temperature.
\appendix\section{}
First, we want to compute the $\lambda\rightarrow 0^+$ asymptotic expansion of  the sum \eqref{eq6_2_1}
 up to the constant term in $\lambda$. Let $\zeta(s;a)$ be the zeta function
\begin{equation}\label{eq5_20_3}
\begin{split}
\zeta(s;a)= \sum_{\boldsymbol{k}\in \Z\times\mathbb{N}^{d-1}}\sum_{l=-\infty}^{\infty}\omega_{\boldsymbol{k}, l}^{-2s}=\frac{1}{\Gamma(s)}\int_0^{\infty} u^{s-1} \sum_{\boldsymbol{k}\in \Z\times\mathbb{N}^{d-1}}\sum_{l=-\infty}^{\infty} \exp\left\{ -u\left(\left( \frac{\pi\left(k_1-\frac{\eta}{2}\right)}{a}\right)^2 +\sum_{i=2}^{\infty}\left(\frac{\pi k_i }{L_i}\right)^2+ (2\pi l T)^2\right)\right\}.
\end{split}
\end{equation}Then
\begin{equation*}
\begin{split}
\Sigma:=\sum_{\boldsymbol{k}\in \Z\times\mathbb{N}^{d-1}}\sum_{l=-\infty}^{\infty}\left( \log \frac{\omega_{\boldsymbol{k}, l}^2}{\mu^2}\right)e^{-\lambda \omega_{\boldsymbol{k}, l}}
=-\left.\frac{\pa}{\pa s}\right|_{s=0}  \frac{ \mu^{2s}}{2\pi i}\int_{c-i\infty}^{c+i\infty}   \Gamma(z) \lambda^{-z}\zeta\left( \frac{z+ 2s}{2};a\right)dz.
\end{split}
\end{equation*}Using the formula
\begin{equation}\label{eq5_20_2}\begin{split}
&\sum_{k_1=-\infty}^{\infty} \exp\left\{-u\left( \frac{\pi\left(k_1-\frac{\eta}{2}\right)}{a}\right)^2\right\}=\frac{a}{\sqrt{\pi u}}\sum_{k_1=-\infty}^{\infty} e^{-\frac{k_1^2 a^2}{u}}e^{\pi i k_1\eta},\end{split}
\end{equation}
one can show that the zeta function $\zeta(s;a)$ \eqref{eq5_20_3}
 only has simple poles at $s=i/2$, $2\leq i\leq d+1$, with residue
$
\text{Res}_{s=\frac{i}{2}}\zeta(s;a) =\frac{c_{d+1-i}(a)}{\Gamma\left(\frac{i}{2}\right)}
$, where \begin{equation*}
c_i(a)=(-1)^i\frac{a}{2^d\pi^{\frac{d+1-i}{2}} T}\sum_{2\leq j_1<j_2<\ldots<j_{d-1-i}\leq d} L_{j_1} \ldots L_{j_{d-1-i}}.
\end{equation*} Therefore,   $\zeta(0;a)=0$ and for $s\in (-1/2, 1/2)$,
\begin{equation*}
\begin{split}
 &\frac{ 1}{2\pi i}\int_{c-i\infty}^{c+i\infty}   \Gamma(z) \lambda^{-z}\zeta\left( \frac{z+ 2s}{2};a\right)dz=2\sum_{i=2}^{d+1}\frac{c_{d+1-i}(a)}{\Gamma\left(\frac{i}{2}\right)}\Gamma(i-2s)\lambda^{2s-i}+\zeta(s;a)+O(\lambda), \;\;\lambda\rightarrow 0^+.
\end{split}
\end{equation*}From this, we find that
\begin{equation}\label{eq5_20_1}
\begin{split}
\Sigma=-2\sum_{i=2}^{d+1}\frac{\Gamma(i)}{\Gamma\left(\frac{i}{2}\right)}c_{d+1-i}(a)\left(\log[\lambda\mu]^2-2\psi(i)\right)\lambda^{-i}-\zeta'(0;a).
\end{split}
\end{equation}

Next, we derive a  formula that is needed  for the computation of the Casimir force \eqref{eq5_20_4}.
Substitute \eqref{eq5_20_2} into \eqref{eq5_20_3}, we find that
\begin{equation*}
\begin{split}
\zeta(s;a) = &\frac{a}{\sqrt{\pi}}\frac{\Gamma\left(s-\frac{1}{2}\right)}{\Gamma(s)}\zeta_{1}\left(s-\frac{1}{2}\right)+\frac{4a}{\sqrt{\pi}\Gamma(s)} \sum_{\boldsymbol{k}\in\mathbb{N}^d}\sum_{l=-i\infty}^{\infty}  \cos(\pi k_1\eta)\left(\frac{k_1 a}{\sqrt{\sum_{i=2}^d\left(\frac{\pi k_i}{L_i}\right)^2+(2\pi l T)^2}}\right)^{s-\frac{1}{2}} \\&\times K_{s-\frac{1}{2}}\left(2k_1 a\sqrt{\sum_{i=2}^d\left(\frac{\pi k_i}{L_i}\right)^2+(2\pi l T)^2}\right),
\end{split}
\end{equation*}where
\begin{equation*}
\zeta_1(s)=\sum_{(k_2,\ldots, k_d)\in \mathbb{N}^{d-1}}\sum_{l=-\infty}^{\infty}\left(\sum_{i=2}^d\left(\frac{\pi k_i}{L_i}\right)^2+(2\pi l T)^2\right)^{-s}.
\end{equation*}Therefore,
\begin{equation}\label{eq5_20_5}
\begin{split}
\frac{\pa}{\pa a}\zeta'(0;a)= -2a\zeta_1\left(-\frac{1}{2}\right)-4 \text{Re}\; \sum_{\boldsymbol{k}\in\mathbb{N}^d}\sum_{l=- \infty}^{\infty} e^{\pi ik_1\eta}  \exp\left(-2k_1a \sqrt{\sum_{i=2}^d\left(\frac{\pi k_i}{L_i}\right)^2+(2\pi l T)^2}\right).
\end{split}
\end{equation}
From this, we obtain the expression for the Casimir force \eqref{eq5_20_6}.

Finally, we list two alternative expressions for the Casimir force $F_{\text{Cas}}^{\infty}(a)$. To study the behavior of the Casimir force when $a\ll 1/T\ll L_i$, $i=2,\ldots, d$, we have
\begin{equation}\label{eq5_27_3}
\begin{split}
& F^{\infty}_{\text{Cas}}(a)= - \frac{\alpha}{2^{d}}\sum_{j=0}^{d-1}(-1)^{d-1-j} \frac{j+1}{\pi^{\frac{j+2}{2}} a^{j+2}} S_j \Gamma\left(\frac{j+2}{2}\right) \sum_{k_1=1}^{\infty}\frac{\cos\left(\pi k_1\eta\right)}{k_1^{j+2}} + \frac{\alpha}{2^{d-1}}\sum_{j=0}^{d-1} \frac{(-1)^{d-1-j}   2^{\frac{j+1}{2}} }{\pi^{\frac{j+1}{2}}}S_j
 \sum_{k_1=-\infty}^{\infty}
\sum_{l=1}^{\infty}\\&\times\frac{\left(\pi \left|k_1-\frac{\eta}{2}\right| \right)^{\frac{j+3}{2}}T^{\frac{j-1}{2}}}{l^{\frac{j-1}{2}}a^{\frac{j+5}{2}}} K_{\frac{j-1}{2}}\left( \frac{\pi l\left|k_1-\frac{\eta}{2}\right|}{aT}\right) + \alpha  T E\left(-\frac{1}{2}; \frac{ \pi}{L_2},\ldots,\frac{\pi}{L_d}\right)- \alpha T \sum_{j=0}^{d-1}  \frac{(-1)^{d-1-j} }{2^{d-j-2}\pi^{\frac{j+2}{2}}}S_j
  \Gamma\left(\frac{j+2}{2}\right)\zeta_R(j+2)T^{j+2}\end{split}\end{equation}\begin{equation}\begin{split}&+
 \frac{\alpha T}{2^{d-2}}\sum_{j=0}^{d-1}(-1)^{d-1-j}  \sum_{2\leq i_1<\ldots<i_j\leq d} \frac{\left[\prod_{p=1}^j L_{i_p}\right]}{\pi^{\frac{j}{2}}}  \sum_{k_1=-\infty}^{\infty}\sum_{(k_{i_1},\ldots, k_{i_j})\in \Z^{j}\setminus\{\boldsymbol{0}\}} \frac{\pi^2\left(k_1-\frac{\eta}{2}\right)^2}{a^3}\left(\frac{\left( \frac{\pi\left(k_1-\frac{\eta}{2}\right)}{a}\right)^2 + (2\pi l T)^2}{\sum_{p=1}^{j}[k_{i_p}L_{i_p}]^2}\right)^{\frac{j-2}{4}}\\&\times  K_{\frac{j-2}{2}}\left( 2\sqrt{\left(\sum_{p=1}^{j}[k_{i_p}L_{i_p}]^2\right)\left(\left( \frac{\pi\left(k_1-\frac{\eta}{2}\right)}{a}\right)^2 + (2\pi l T)^2\right)}\right)-\frac{\alpha T}{2^{d-2}}\sum_{j=1}^{d-1}(-1)^{d-1-j}  \sum_{2\leq i_1<\ldots<i_j\leq d}\frac{\left[\prod_{p=1}^j L_{i_p}\right]}{\pi^{\frac{j+1}{2}}}\\&\times \sum_{l=1}^{\infty}\sum_{(k_{i_1},\ldots, k_{i_j})\in \Z^{j}\setminus\{\boldsymbol{0}\}} \left(\frac{2\pi lT}{\sqrt{ \sum_{p=1}^{j}[k_{i_p}L_{i_p}]^2}}\right)^{\frac{j+1}{2}}K_{\frac{j+1}{2}}\left( 4\pi l T\sqrt{ \sum_{p=1}^{j}[k_{i_p}L_{i_p}]^2}\right),
\end{split}
\end{equation} where
  $E\left(-\frac{1}{2}; \frac{ 1}{L_2},\ldots,\frac{1}{L_d}\right)$ is the analytic continuation of the function
\begin{equation*}\begin{split}
&E\left(s; \frac{\pi}{L_2},\ldots,\frac{\pi}{L_d}\right) =\sum_{(k_2,\ldots, k_d)\in \mathbb{N}^{d-1}}\left( \left[\frac{\pi k_2}{L_2}\right]^2+\ldots +  \left[\frac{\pi k_d}{L_d}\right]^2\right)^{-s}\end{split}
\end{equation*}to $s=-1/2$, which is equal to the zeta regularized Casimir energy for massless scalar field inside a rectangular box   $[0,L_2]\times\ldots \times [0,L_d]$ subject to Dirichlet boundary conditions (see e.g. \cite{5_27_1, 5_27_2, 5_27_3}). It can be shown that (see e.g. \cite{5_27_4, 5_27_5})
\begin{equation}\label{eq5_27_5}
\begin{split}
&E\left(-\frac{1}{2}; \frac{\pi}{L_2},\ldots,\frac{\pi}{L_d}\right)=-\frac{1}{2^{d-1}}\sum_{j=1}^{d-1} (-1)^{d-1-j}\frac{\Gamma\left(\frac{j+1}{2}\right)  }{\pi^{\frac{j+1}{2}}L_d^j}   \zeta_R(j+1)\sum_{2\leq i_1<\ldots<i_{j-1}}\left[\prod_{p=1}^{j-1}L_{i_p}\right]-\frac{1}{2^{d-2}}\sum_{j=2}^{d-1}(-1)^{d-1-j} \\
&\times\sum_{2\leq i_1<\ldots<i_{j-1}\leq d-1}\left[\prod_{p=1}^{j-1}L_{i_p}\right]\sum_{(k_{i_1},\ldots, k_{i_{j-1}})\in \Z^{j-1}\setminus\{\mathbf{0}\}}\sum_{k_d=1}^{\infty}  \left(\frac{k_d}{L_d \sqrt{\sum_{p=1}^{j-1}\left(k_{i_p}L_{i_p}\right)^2}}\right)^{\frac{j}{2}} K_{\frac{j}{2}}\left(\frac{2\pi k_d}{L_d}\sqrt{ \sum_{p=1}^{j-1}\left(k_{i_p}L_{i_p}\right)^2}\right).
\end{split}
\end{equation}

To study the behavior of the Casimir force when $  1/T\ll a\ll L_i$, $i=2,\ldots, d$, we have
\begin{equation}\label{eq5_27_4}
\begin{split}
& F^{\infty}_{\text{Cas}}(a)=
-\frac{\alpha T}{2^{d-1}}\sum_{j=0}^{d-1}   \frac{(-1)^{d-1-j}}{\pi^{\frac{j+1}{2}}}\frac{j S_j }{a^{j+1}}\Gamma\left(\frac{j+1}{2}\right) \sum_{k_1=1}^{\infty}\frac{\cos(\pi k_1\eta)}{k_1^{j+1}}- \frac{\alpha T}{2^{d-3}}\sum_{j=0}^{d-1}    \frac{(-1)^{d-1-j}}{\pi^{\frac{j+1}{2}}}j S_j\sum_{k_1=1}^{\infty}\sum_{l=1}^{\infty}\cos(\pi k_1\eta) \\&\times\left(\frac{2\pi l T}{k_1 a}\right)^{\frac{j+1}{2}} K_{\frac{j+1}{2}}\left( 4\pi k_1 l Ta\right)- \frac{\alpha T}{2^{d-4}}\sum_{j=0}^{d-1}  \frac{(-1)^{d-1-j}}{\pi^{\frac{j+1}{2}}} S_j \sum_{k_1=1}^{\infty}\sum_{l=1}^{\infty}\cos(\pi k_1\eta)\frac{\left(2\pi l T\right)^{\frac{j+3}{2}}}{\left(k_1 a\right)^{\frac{j-1}{2}}}K_{\frac{j-1}{2}}\left( 4\pi k_1 l Ta\right) \\&+
 \frac{\alpha T}{2^{d-2}}\sum_{j=0}^{d-1}(-1)^{d-1-j}  \sum_{2\leq i_1<\ldots<i_j\leq d} \frac{\left[\prod_{p=1}^j L_{i_p}\right]}{\pi^{\frac{j}{2}}}  \sum_{k_1=-\infty}^{\infty} \sum_{(k_{i_1},\ldots, k_{i_j})\in \Z^{j}\setminus\{\boldsymbol{0}\}}\frac{\pi^2\left(k_1-\frac{\eta}{2}\right)^2}{a^3} \left(\frac{\left( \frac{\pi\left(k_1-\frac{\eta}{2}\right)}{a}\right)^2 + (2\pi l T)^2}{\sum_{p=1}^{j}[k_{i_p}L_{i_p}]^2}\right)^{\frac{j-2}{4}}\\&\times K_{\frac{j-2}{2}}\left( 2\sqrt{\left(\sum_{p=1}^{j}[k_{i_p}L_{i_p}]^2\right)\left(\left( \frac{\pi\left(k_1-\frac{\eta}{2}\right)}{a}\right)^2 + (2\pi l T)^2\right)}\right) + \alpha T E\left(-\frac{1}{2}; \frac{\pi}{L_2},\ldots,\frac{\pi}{L_d}\right) -\frac{\alpha T}{2^{d-2}}\sum_{j=1}^{d-1}(-1)^{d-1-j}\\&\times  \sum_{2\leq i_1<\ldots<i_j\leq d}\frac{\left[\prod_{l=1}^j L_{i_l}\right]}{\pi^{\frac{j+1}{2}}}\sum_{l=1}^{\infty}\sum_{(k_{i_1},\ldots, k_{i_j})\in \Z^{j}\setminus\{\boldsymbol{0}\}} \left(\frac{2\pi lT}{\sqrt{ \sum_{p=1}^{j}[k_{i_p}L_{i_p}]^2}}\right)^{\frac{j+1}{2}}K_{\frac{j+1}{2}}\left( 4\pi l T\sqrt{ \sum_{p=1}^{j}[k_{i_p}L_{i_p}]^2}\right).
\end{split}
\end{equation}
\begin{acknowledgments}
This project is   funded by Ministry of Science, Technology and Innovation, Malaysia under e-Science fund 06-02-01-SF0080.
\end{acknowledgments}

\end{document}